\documentclass
[superscriptaddress,secnumarabic,amssymb,amsmath,nobibnotes,aps,prd,showkeys,showpacs,nofootinbib,onecolumn]{revtex4}%
\usepackage{graphicx}
\usepackage{epsf}
\usepackage{bm}
\usepackage{amsmath}
\usepackage{amsfonts}
\usepackage{amssymb}
\usepackage{epstopdf}%
\providecommand{\U}[1]{\protect\rule{.1in}{.1in}}
\newcommand{\be}{\begin{equation}}
\newcommand{\ee}{\end{equation}}

\newcommand{\mincir}{\raise
-3.truept\hbox{\rlap{\hbox{$\sim$}}\raise4.truept\hbox{$<$}\ }}
\newcommand{\magcir}{\raise
-3.truept\hbox{\rlap{\hbox{$\sim$}}\raise4.truept\hbox{$>$}\ }}

\begin{document}
\title{Duality transformation and conformal equivalent scalar-tensor theories}
\author{Gabriele Gionti, S.J.}
\email{ggionti@specola.va}
\affiliation{Specola Vaticana, V-00120 Vatican City, Vatican City State, Vatican
Observatory Research Group, Steward Observatory, The University Of Arizona,
933 North Cherry Avenue, Tucson, Arizona 85721, USA}
\affiliation{INFN, Laboratori Nazionali di Frascati, Via E. Fermi 40, 00044 Frascati, Italy}
\author{Andronikos Paliathanasis}
\email{anpaliat@phys.uoa.gr}
\affiliation{Instituto de Ciencias F\'{\i}sicas y Matem\'{a}ticas, Universidad Austral de
Chile, Valdivia, Chile}
\affiliation{Institute of Systems Science, Durban University of Technology, PO Box 1334,
Durban 4000, Republic of South Africa}

\begin{abstract}
We deal with the duality symmetry of the Dilaton field in cosmology and
specifically with the so-called Gasperini-Veneziano duality transformation. In
particular, we determine two conformal equivalent theories to the Dilaton
field, and we show that under conformal transformations Gasperini-Veneziano
duality symmetry does not survive. Moreover, we show that those theories share
a common conservation law, of Noetherian kind, while the symmetry vector which
generates the conservation law is an isometry only for the Dilaton field.
Finally, we show that the Lagrangian of the Dilaton field is equivalent with
that of the two-dimensional \textquotedblright hyperbolic
oscillator\textquotedblright\ in a Lorentzian space whose $O(d,d)$ invariance
is transformed to the Gasperini-Veneziano duality invariance in the original coordinates.

\end{abstract}
\maketitle
\date{\today$\frac{{}}{{}}$}

\section{Introduction}

Theories of gravity which extend General Relativity have drawn the attention
of the scientific society in that last decade because they provide geometric
methods to solve the problem of the late-time acceleration of the universe
without the necessity to consider dark energy, for instance see
\cite{faraonibook,re1,re2,re3,re4,re5}. \ On the other hand, an important
property of the two-dimensional conformal field theory is the duality symmetry
which has important consequences in cosmology
\cite{newref,Alvarez,review1,ref1,ref2,ref3,ref4}. The duality symmetry is
characterized by the invariance of the action integral, therefore the
corresponding Euler-Lagrange equations, under this transformation, remain the
same. Usually, in String Theory, when the \textquotedblleft
radius\textquotedblright\ $\mathcal{R}$~of the compactified geometry changes
such as $\bar{\mathcal{R}}\rightarrow\mathcal{R}^{-1}$. Duality transformation
is a discrete transformation and an isometry should exist for the underlying
manifold \cite{bush1,bush2}.

If one considers the tree level effective action of gravity minimally coupled
with the Dilaton field, derived by imposing, at one loop, the conformal
invariance of the sigma model at quantum level \cite{bush1,bush2}, then they get%

\begin{equation}
S=\int d^{4}x\sqrt{-G}e^{-2\phi}\left(  R-4G^{\mu\nu}\phi_{;\mu}\phi_{;\nu
}+\Lambda-\frac{1}{12}H_{\mu\nu\rho}H^{\mu\nu\rho}\right)  . \label{total}%
\end{equation}
Here $\Lambda$ is the cosmological constant, $\phi$ the Dilaton field,
$H_{\mu\nu\rho}$ is expressed in terms of $B_{\mu\nu}$ as $H_{\mu\nu\rho
}=\partial_{\mu}B_{\nu\rho}+cyclic\;permutations$. In case $G$ , $B$ and
$\phi$ are functions of the time only \cite{mesh,mesh2}, one can prove that%

\begin{equation}
G=
\begin{pmatrix}
-1 & 0\\
0 & G(t)
\end{pmatrix}
, B=
\begin{pmatrix}
0 & 0\\
0 & B(t)
\end{pmatrix}
\end{equation}

It has been shown in \cite{gasperini} that introducing a $2d\times2d$ matrix
and a field $\Psi$%

\begin{equation}
M=%
\begin{pmatrix}
G^{-1} & -G^{-1}B\\
BG^{-1} & G-BG^{-1}B
\end{pmatrix}
,~\Psi=2\phi-ln\sqrt{detG}%
\end{equation}
with these positions the previous action (\ref{total}) becomes%

\begin{equation}
S=\int dte^{-\Psi}\big (\Lambda+({\dot{\Psi}})^{2}+\frac{1}{8}Tr\left[
{\dot{M}}\eta{\dot{M}}\eta\right]  \big ) \label{modi}%
\end{equation}
in which $\eta_{\sigma\lambda}$ is the $O(d,d)$ off diagonal metric defined as%

\begin{equation}
\eta_{\sigma\lambda}=%
\begin{pmatrix}
0 & 1\\
1 & 0
\end{pmatrix}
.
\end{equation}
Furthermore, it is straightforward to see that action (\ref{modi}) is $O(d,d)$
invariant, that is, $\Psi\mapsto\Psi,~M\mapsto\Omega^{T}M\Omega,$ where
$\Omega^{T}\eta\Omega=\eta$. In the case $B=0$ and $\Omega=\eta$, one gets the
\textquotedblleft reflection\textquotedblright\ of the $O(d,d)$ group which is
formulated as follows%

\begin{equation}
G(t)\mapsto G(t)^{-1}\;\;,\;\phi\mapsto\phi-ln\sqrt{detG} \label{trasformo}%
\end{equation}
so that considering a spatially flat Friedmann-Lema\^{\i}tre-Robertson-Walker
metric (FLRW) with scale factor $a(t)$,~and $\sqrt{-g}=a^{3}$, the $O(d,d)$
transformation becomes the Gasperini-Veneziano duality
transformation~\thinspace$a\mapsto a^{-1}$ and $\phi\mapsto\phi-3a.~$

Indeed, the Gasperini-Veneziano duality transformation maps solutions into
solutions, and being generated, in general, by a continuous group, $O(d,d)$
its is associated with an isometry of the Lagrangian, for a review see
\cite{revven}. This means that there exists a Vector field $X$ such that the
Lie derivative $L_{X}$ of the Lagrangian $\mathcal{L}$ is zero $L_{X}%
\mathcal{L}=0,~$which means that $X$ is an isometry for the Lagrangian.~That
is an important property for the existence of a duality transformation and is
used later to explain how duality symmetry is lost under conformal
transformations, for conformal equivalent theories.

Recently in \cite{Cap01} a similar transformation to that of the
Gasperini-Veneziano duality transformation \cite{gasperini2} was studied and
it's connection with point symmetries in the case of the modified
gravitational fourth-order theory, the $f\left(  R\right)  $-gravity, while a
study on the relation between continuous and discrete symmetries in
gravitational theories recently performed in \cite{ancap}.

In this letter we revise the problem and in our approach we consider the
property, that scalar-tensor theories and $f\left(  R\right)  $-gravity are
conformal equivalent. We show that because of this constraint those conformal
equivalent theories share a common invariant transformation for the Lagrangian
in which only for the Dilaton field, that transformation reduce to the duality
Gasperini-Veneziano duality transformation.

\section{Conformal equivalent theories and symmetries}

For the convenience of the reader we briefly discuss the conformal equivalence
of scalar-tensor theories. \ Let $\varphi$ be a minimally coupled scalar field
in which the field equations are given by the variation of the action
Integral
\begin{equation}
S_{\varphi}=\int d^{4}x\sqrt{-g}\left(  F\left(  \varphi\right)  R\left(
g\right)  -\frac{1}{2}g^{\mu\nu}\varphi_{;\mu}\varphi_{;\nu}+V\left(
\varphi\right)  \right)  \label{lan.01}%
\end{equation}
where $g_{\mu\nu}$ is the metric tensor for a four-dimensional Riemannian
space of Lorentzian signature, and Ricciscalar $R\left(  g\right)  $.

Consider the conformal equivalent metric $g_{\mu\nu}=N^{2}\left(
x^{k}\right)  \gamma_{\mu\nu},~$where now the Ricciscalars $R\left(
\gamma\right)  ,~$and $R\left(  g\right)  $ are related as follows
\cite{hawellis}%
\begin{equation}
R\left(  g\right)  =N^{-2}R\left(  \gamma\right)  -6N^{-3}\gamma^{\mu\nu
}N_{;\mu\nu}. \label{lan.02}%
\end{equation}
By replacing (\ref{lan.02}) in (\ref{lan.01}) it follows%
\begin{equation}
S_{\varphi}=\int d^{4}x\sqrt{-\gamma}\left(  F\left(  \varphi\right)
N^{2}R\left(  \gamma\right)  -\frac{1}{2}N^{2}\gamma^{\mu\nu}\varphi_{;\mu
}\varphi_{;\nu}+N^{4}V\left(  \varphi\right)  -6F\left(  \varphi\right)
N\gamma^{\mu\nu}N_{;\mu\nu}\right)  \label{lan.03}%
\end{equation}
Integration by parts in the last term of expression (\ref{lan.03}) gives%
\begin{equation}
A_{0}=\int d^{4}x\sqrt{-\gamma}\left(  6NF_{,\varphi}\gamma^{\mu\nu}N_{;\mu
}\phi_{;\nu}+6F\left(  \varphi\right)  \gamma^{\mu\nu}N_{;\mu}N_{;\nu}\right)
\label{lan.04}%
\end{equation}
Function $N\left(  x^{k}\right)  $ is not a real degree of freedom and without
loss of generality can be neglected by assuming that $N\left(  x^{k}\right)
=N\left(  \varphi\left(  x^{k}\right)  \right)  .~$Hence with that
consideration the Action Integral (\ref{lan.03}) becomes%
\begin{equation}
S_{\varphi}=\int d^{4}x\sqrt{-\gamma}\left(  F\left(  \varphi\right)
N^{2}R\left(  \gamma\right)  -\frac{1}{2}W\left(  \varphi\right)  \gamma
^{\mu\nu}\varphi_{;\mu}\varphi_{;\nu}+\hat{V}\left(  \varphi\right)  \right)
\label{lan.005}%
\end{equation}
where
\begin{equation}
W\left(  \varphi\right)  =-2\left(  -\frac{1}{2}N^{2}+6NF_{,\varphi
}N_{,\varphi}+6F\left(  \varphi\right)  \left(  N_{,\varphi}\right)
^{2}\right)  ~,~\hat{F}\left(  \varphi\right)  =F\left(  \varphi\right)
N^{2},~\text{and,~}\hat{V}\left(  \varphi\right)  =N^{4}V\left(
\varphi\right)  . \label{lan.006}%
\end{equation}

The action integrals, (\ref{lan.01}), and (\ref{lan.005}), are different but
they describe two conformal equivalent scalar-tensor theories. Moreover, in
order to make clear that actions (\ref{lan.01}), and (\ref{lan.005}) describe
different fields, we define the new field $\Phi$ such that $\Phi=\int
\sqrt{W\left(  \varphi\right)  }d\varphi$, in which (\ref{lan.005}) takes the
simplest form~\cite{faraonibook}%
\begin{equation}
S_{\Phi}=\int d^{4}x\sqrt{-\gamma}\left(  \hat{F}\left(  \Phi\right)
R-\frac{1}{2}\gamma^{\mu\nu}\Phi_{;\mu}\Phi_{;\nu}+\hat{V}\left(  \Phi\right)
\right)  . \label{lan.007}%
\end{equation}
It is important to mention here that when $\hat{F}\left(  \Phi\right)
=const.$, that is $F\left(  \varphi\right)  N\left(  \varphi\right)
^{2}=const.$, the action integral (\ref{lan.007}) describes a minimally
coupled scalar field model.

Consider now the Dilaton field with action integral (\ref{total}) and
$H_{\mu\nu\rho}=0$, that is,
\begin{equation}
S_{\left(  Dilaton\right)  }=\int d^{4}x\sqrt{-g}e^{-2\phi}\left(
R-4g^{\mu\nu}\phi_{;\mu}\phi_{;\nu}+\Lambda\right)  . \label{lan.008}%
\end{equation}
which without loss of generality can be written in the form of the Action
(\ref{lan.01}) with $F\left(  \varphi\right)  =\frac{1}{8}\varphi^{2},$
and~$V\left(  \phi\right)  =\frac{\Lambda}{8}\varphi^{2}$, while the two
fields $\phi,~$and $\varphi$ are related as $2\sqrt{2}e^{-\phi}=\varphi.$
Consequently, from the above formulas we find that the minimally coupled
scalar field equivalent Action is
\begin{equation}
S_{A}=\int d^{4}x\sqrt{-\gamma}\left(  R-\frac{1}{2}\gamma^{\mu\nu}\Phi_{;\mu
}\Phi_{;\nu}+\hat{\Lambda}e^{-\lambda\Phi}\right)  . \label{lan.009}%
\end{equation}

\noindent in which $\hat{\Lambda}=64\Lambda$ and $\lambda=\frac{1}{\sqrt{5}}$.
However, from our discussion now it is clear that the later action is not
unique and there exists a family of conformal equivalent actions with the
Dilaton field (\ref{lan.008}). For instance if we perform the (second)
conformal transformation $\gamma_{\mu\nu}=e^{\frac{\sqrt{3}}{3}\Phi}%
\kappa_{\mu\nu}~,~\Phi=\sqrt{3}\ln\sigma,~~$in (\ref{lan.009}), we find the
conformal equivalent action
\begin{equation}
S_{B}=\int d^{4}x\sqrt{-\gamma}\left(  \sigma R+\hat{\Lambda}\sigma^{\mu
}\right)  ~,~\mu=\left(  2-\frac{\sqrt{3}}{\sqrt{5}}\right)  \label{lan.010}%
\end{equation}
which is the O'Hanlon action with a power-law potential $V\left(
\sigma\right)  \simeq\sigma^{\mu}~$\cite{hanlon}.

It is well-known that O'Hanlon theory is equivalent with $f\left(  R\right)
$-gravity~\cite{Buda}~with the use of Lagrange multiplier \cite{so1,so2}.
Specifically, it is an alternative way to write the action of $f\left(
R\right)  $-gravity, as a special case of Hordenski theory \cite{hor1}. The
potential of the O'Hanlon theory and the $f\left(  R\right)  $-gravity are
related by the Clairaut equation~$f_{,R}\left(  R\right)  R-f\left(  R\right)
=V\left(  f_{,R}\right)  ,~$where for the power-law potential in
(\ref{lan.010}) the $f\left(  R\right)  $ equivalent theory is derived to be
the power-law $f\left(  R\right)  $~model.%

\begin{equation}
f\left(  R\right)  \simeq R^{\frac{\mu}{\mu-1}}\text{~,~}\mu=\left(
2-\frac{\sqrt{3}}{\sqrt{5}}\right)  . \label{lan.011}%
\end{equation}

Now consider the underlying space to be that of FLRW geometry with zero
spatial curvature, that is%
\begin{equation}
ds^{2}=-N^{2}\left(  t\right)  dt^{2}+a^{2}\left(  t\right)  \left(
dx^{2}+dy^{2}+dz^{2}\right)  \label{lan.012}%
\end{equation}
where $a\left(  t\right)  $ is the scale factor and $N\left(  t\right)  $ is
the lapse function. We continue by assuming that the matter source inherits
the spacetime symmetries.

For the three conformal equivalent actions\footnote{It is important to mention
that conformal equivalent theories are not necessary and physical equivalent,
for a discussion see \cite{farval00}.}, (\ref{lan.008}), (\ref{lan.009}), and
(\ref{lan.010}) for the Dilaton field $\varphi,~$the minimally coupled field
$\Phi$, and the $f\left(  R\right)  $ gravity (field $\sigma$),~respectively;
the field equations in the spatially flat FLRW background space (\ref{lan.012}%
) are described by the following point-like Lagrangians%
\begin{equation}
L_{\varphi}\left(  a,\dot{a},\varphi,\dot{\varphi}\right)  =\frac{1}{N}\left(
6a\varphi^{2}\dot{a}^{2}+12a^{2}\varphi\dot{a}\dot{\varphi}-4a^{3}\dot
{\varphi}^{2}\right)  -\Lambda Na^{3}\varphi^{2}, \label{lan.013}%
\end{equation}%
\begin{equation}
L_{\Phi}\left(  a,\dot{a},\Phi,\dot{\Phi}\right)  =\frac{1}{N}\left(
3a\dot{a}^{2}-\frac{1}{2}a^{3}\dot{\Phi}^{2}\right)  +\hat{\Lambda
}Ne^{-\lambda\Phi}, \label{lan.014}%
\end{equation}
and%
\begin{equation}
L_{\sigma}\left(  a,\dot{a},\sigma,\dot{\sigma}\right)  =\frac{1}{N}\left(
6a\sigma\dot{a}^{2}+6a^{2}\dot{a}\dot{\sigma}\right)  +\hat{\Lambda}%
N\sigma^{\mu}. \label{lan.015}%
\end{equation}
Those three point-like Lagrangians are singular, and are conformally related
because of the constraint condition $\mathcal{H}\equiv\frac{\partial
L}{\partial N}=0$. Without loss of generality in the following we set
$N\left(  t\right)  =1$.

We observe that only Dilaton's Lagrangian (\ref{lan.013}) is invariant under
the Gasperini-Veneziano duality transformation, which on the coordinates
$\left\{  a,\varphi\right\}  $ is $a\rightarrow a^{-1}~,~\varphi
\rightarrow\varphi a^{3}.~$\ The three point-like Lagrangians share another
property, that is, all admit linear in the momentum conservation laws which
are generated by the same conformal killing vector field of the two
dimensional minisuperspace\footnote{For more details on how those conservation
laws are constructed and related, we refer the reader to
\cite{paper1,paper2,paper3,paper4,paper5} and references therein.} (but in
different coordinates).

The conservation laws are
\begin{equation}
I_{\varphi}\left(  a,\dot{a},\varphi,\dot{\varphi}\right)  =a^{2}\varphi
^{2}\dot{a}-4a^{3}\varphi\dot{\varphi}, \label{lan.017}%
\end{equation}%
\begin{equation}
I_{\Phi}\left(  a,\dot{a},\Phi,\dot{\Phi}\right)  =a^{2}\left(  \lambda\dot
{a}-a\dot{\Phi}\right)  , \label{lan.018}%
\end{equation}%
\begin{equation}
I_{\sigma}\left(  a,\dot{a},\sigma,\dot{\sigma}\right)  =2a^{2}\sigma\left(
\mu-2\right)  \dot{a}+a^{3}\left(  \mu+1\right)  \dot{\sigma} \label{lan.019}%
\end{equation}
while the corresponding generators of the conservation laws are%
\begin{equation}
X_{\varphi}=a\partial_{a}-\frac{3}{2}\varphi\partial_{\varphi}~,~X_{\Phi
}=a\partial_{a}+\frac{6}{\lambda}\partial_{\Phi}~,~X_{\sigma}=\frac{\left(
\mu+1\right)  }{6}a\partial_{a}+\sigma\partial_{\sigma}, \label{lan.020}%
\end{equation}
which are Noether (point) symmetries for the Lagrangians (\ref{lan.013}),
(\ref{lan.014}), and (\ref{lan.015}) respectively.

However,~functions $I_{\Phi},~I_{\sigma}$ are \textquotedblleft
weak\textquotedblright\ conservation laws in the sense that someone has to
impose the constraint condition $\mathcal{H}=0$, in order $\frac{dI}{dt}=0$,
and that is because it is held $\frac{dI_{\Phi}}{dt}=2\mathcal{H}_{\Phi}$ and
$\frac{dI_{\sigma}}{dt}=2\mathcal{H}_{\sigma}$. Hence, vector fields $X_{\Phi
}$, and $X_{\sigma}$ are Homothetic symmetries for the minisuperspace of
Lagrangians $L_{\Phi}$, and $L_{\sigma}$, and only $X_{\varphi}$ is a Killing
symmetry, for $L_{\varphi},$ which is a necessity condition in order
Gasperini-Veneziano duality transformation to exist. \ Therefore, we can say
that duality invariance is lost under conformal transformations and the
power-law $f\left(  R\right)  $-gravity does not admit any duality symmetry,
for arbitrary value $\mu$,~such that $\mu\neq0,1$.

\subsection{Classical mechanical analogue}

In order to find the classical mechanical analogue of the Gasperini-Veneziano
duality transformation we perform the change of variables
\begin{equation}
a=u^{\frac{2}{7}-\frac{\sqrt{15}}{21}}v^{\frac{2}{7}+\frac{\sqrt{15}}{21}%
}~,~\varphi=u^{\frac{1}{14}+\frac{\sqrt{15}}{14}}v^{\frac{1}{14}-\frac
{\sqrt{15}}{14}} \label{lan.021}%
\end{equation}
in (\ref{lan.013}) where $u=x+y~$and $v=x-y$. Because that is a coordinate
transformation the new Lagrangian is equivalent with the original without
imposing any constraint \cite{arnoldbook}. \ The resulting Lagrangian after
the coordinate transformation (\ref{lan.021}) is%
\begin{equation}
L_{\varphi}\left(  x,\dot{x},y,\dot{y}\right)  =\frac{1}{2}\left(  \dot{x}%
^{2}-\dot{y}^{2}\right)  +\bar{\Lambda}\left(  x^{2}-y^{2}\right)
\end{equation}
which describes the hyperbolic \textquotedblleft oscillator\textquotedblright%
\ in a two-dimensional Riemannian manifold of Lorentzian signature. The field
equations are well known that are maximally symmetric
\cite{osci1,osci2,osci3,osci4} and the rotation symmetry of the $M^{2}$ space
provides the discrete dual symmetry $x\rightarrow y$, $y\rightarrow x$, which
in the original coordinates of $\left\{  a,\phi\right\}  $ becomes the
Gasperini-Veneziano duality symmetry. That is an alternative way to prove that
the original of the Gasperini-Veneziano duality transformation is the
$O\left(  d,d\right)  $ symmetry.

\section{Conclusions}

In this letter, we studied the Gasperini-Veneziano duality symmetry for the
Dilaton field for conformal equivalent theories. We found that duality
symmetry does not survive under conformal transformations and we show that
while all the conformal equivalent theories admit a conservation law of the
same origin, only in the frame of the Dilaton field the conservation law
follows from an isometry of the Lagrangian. Furthermore, we wrote the
Lagrangian of the Dilaton field in it's classical mechanical analogue of the
two-dimensional hyperbolic \textquotedblleft oscillator\textquotedblright%
\ where the duality symmetry becomes the rotation symmetry, the so-called
$O\left(  d,d\right)  $ symmetry.

Finally, in the context of $f\left(  R\right)  $-gravity, we show that there
is not any duality symmetry for the power-law theory while the analysis which
we presented in this work can be applied in extendent theories to determine
discrete symmetries.

\begin{acknowledgments}
AP was financially supported by FONDECYT grants 3160121 and want to thank Guy
Consolmagno, S.J. and the members of the Vatican Observatory (Specola
Vaticana) for the invitation and the hospitality provided while this work was
carried out.
\end{acknowledgments}

\end{document}